\documentclass[fleqn,usenatbib]{mnras}
\usepackage{newtxtext,newtxmath}

\usepackage[T1]{fontenc}
\DeclareRobustCommand{\VAN}[3]{#2}
\let\VANthebibliography\thebibliography
\def\thebibliography{\DeclareRobustCommand{\VAN}[3]{##3}\VANthebibliography}

\usepackage{graphicx}	
\usepackage{amsmath}	
\usepackage{xcolor}

\newcommand{\galform}{\textsc{galform}}

\newcommand{\Color}{\textsc{color}}
\newcommand{\PGadget}{\textsc{P-Gadget-3}}

\newcommand{\Subfind}{\textsc{subfind}}

\title[Failed MWs]{The progenitor galaxies of stellar haloes as ``failed'' Milky Ways}

\author[S. Bose \& A. Deason]{
Sownak Bose$^{1}$\thanks{E-mail: sownak.bose@durham.ac.uk (SB)} and
Alis J. Deason$^{1}$\thanks{E-mail: alis.j.deason@durham.ac.uk (AD)}
\\
$^{1}$Institute for Computational Cosmology, Department of Physics, Durham University, Durham DH1 3LE, UK
}

\date{Accepted XXX. Received YYY; in original form ZZZ}

\pubyear{2022}

\begin{document}
\label{firstpage}
\pagerange{\pageref{firstpage}--\pageref{lastpage}}
\maketitle

\begin{abstract}
The stellar halo of the Milky Way records the history of its interactions with dwarf galaxies, whose subsequent destruction results in the formation of an extended stellar component. Recent works have suggested that galaxies with masses comparable to the Large Magellanic Cloud (LMC, $M_\star \sim 10^9\,{\rm M}_\odot$) may be the primary building blocks of the stellar halo of our Galaxy. We use cosmological simulations of the $\Lambda$ Cold Dark Matter model to investigate LMC-mass galaxies at $z=1-2$ using a semi-analytic model of galaxy formation. We find that LMC analogues at $z=2$ evolve until the present day along three distinct pathways: (1) those that are destroyed in Milky Way-mass hosts; (2) those that are themselves the main progenitors of Milky Way-mass galaxies; and (3) those that survive until $z=0$, with stellar mass $\sim$1.0~dex lower than typical Milky Ways. We predict that the properties of these galaxies at $z=2$ (stellar metallicities, sizes, gas content etc.) are largely indistinguishable, irrespective of which of these pathways is eventually taken; a survey targeting such galaxies in this redshift range would struggle to tell apart a `destroyed' stellar halo progenitor from a `surviving' LMC analogue. The only factor that determines the eventual fate of these galaxies is their proximity to a neighbouring Milky Way main progenitor at $z=2$: while the mean separation to a `surviving' galaxy is around 7 Mpc, it is only 670 kpc to a `destroyed' galaxy. This suggests that old stellar populations in the Milky Way may share intrinsic (i.e. non-dynamical) properties that are essentially indistinguishable from progenitors of its stellar halo. 
\end{abstract}

\begin{keywords}
Galaxy: formation -- galaxies: dwarf -- {\it (galaxies:)} Local Group -- methods: numerical
\end{keywords}



\section{Introduction}
\label{sec:intro}

The history of the Milky Way's formation is a tumultuous one: it is punctuated by the merger and accretion of smaller, dwarf galaxies, each merger changing the mass, kinematics and composition of our Galaxy. Perhaps the most important record-keeper of these events is the so-called stellar halo, extending out to 100s of kiloparsecs from the Galactic centre \citep[]{Deason2020}. While only a small fraction ($\lesssim1\%$) of the total stellar mass is contained within the stellar halo \citep[][]{Carney1989,Helmi2008,Deason2019,Mackereth2020}, the stellar populations comprising it are the remnants of ancient dwarf galaxies that were accreted into the gravitational potential of the Milky Way's dark matter halo. A stellar halo, therefore, acts as a fossil record for the demographics and the properties of dwarf galaxies at high redshift \citep[see, e.g.,][for a characterisation of some of the substructures building the stellar halo of the Milky Way]{Naidu2020}. The very existence of stellar haloes is a fundamental prediction of hierarchical theories of structure formation \citep[e.g.][]{Helmi1999,Bullock2005,Cooper2010}, and their presence (or not) in galaxies across a wide mass range may reveal important clues regarding structure formation on small-scales, such as the merger history and modes of feedback in dwarf galaxies \citep{Fitts2018,Martin2021,Tarumi2021,KadoFong2022}, and perhaps even the nature of the dark matter \citep{Deason2022}.

The synthesis of high precision observations of our own Galaxy enabled by surveys like {\it Gaia}, APOGEE, and SDSS, combined with modern generations of cosmological simulations, have helped develop an increasingly consistent picture of the assembly history of the Milky Way. Amongst these events, perhaps the most readily-observed examples are that of the Small and Large Magellanic Clouds (SMC and LMC), which are thought to have been accreted into the potential of the Milky Way's halo 2-3 Gyrs ago and which are now on their first orbit in the Galactic potential \citep[e.g.][]{Kallivayalil2006,Besla2007,Sales2011,BoylanKolchin2011,Besla2015,Shao2018}. The archaeology of metal-rich halo stars identified in {\it Gaia} data has provided evidence of yet another tantalising event in the past history of our Galaxy: an ancient dwarf galaxy merger, similar in mass to the LMC, between 8 and 11 Gyrs ago, roughly around the time of the formation of the Galactic disc
\citep{Belokurov2018,Helmi2018,Myeong2018,Gallart2019}. This merger event, known as the {\it Gaia}-Enceladus-Sausage not only provides a pointer to the early accretion history of the Galactic halo that shaped the subsequent growth of the Milky Way, but may have also altered the census of satellite galaxies that orbit its potential at present day \citep[e.g.][]{Bose2020}.

While we expect there to be several merger events that have contributed to the overall structure of the stellar halo as we observe it today \citep[e.g.][]{Kruijssen2020,Forbes2020}, it has been suggested that the properties of the Milky Way's stellar halo are, to first order, set largely by the properties of the most massive ancient dwarf galaxy merger it experiences in its lifetime \citep[see, e.g.,][]{Deason2016,DSouza2018}. There is also evidence from the age-metallicity distribution of globular clusters in the Milky Way \citep{Kruijssen2019}, as well as from metal-poor stars near the Galactic Centre \citep{Horta2021} that the most massive ancient merger was with a galaxy in the mass range $5\times10^8-2\times10^9\,{\rm M}_\odot$, sometimes dubbed the ``Kraken''. While evidence for such an event is under debate \citep[e.g.][]{Belokurov2022,Orkney2022}, it is not too much of a stretch to state that the most significant dwarf galaxy merger in the Milky Way corresponds to the accretion of a galaxy with mass comparable to that of the LMC. It is interesting, therefore, to consider the properties of these progenitors of the stellar halo at high redshift -- during an epoch in which they were isolated galaxies in their own right, long before being destroyed within the Galactic halo. Are there any identifiable physical properties that set these galaxies apart from the remainder of the population at high redshift? If so, could they be distinguished using multi-wavelength observations at these redshifts in a way that allows us to separate a galaxy that is about to be ``destroyed'' (i.e., become a stellar halo progenitor) from one that ``survives''? These questions are particularly pertinent in the era of the {\it James Webb Space Telescope} (JWST) which, with its NIRCam instrument, may be capable of identifying these progenitor galaxies at $z\gtrsim1$ \citep{Evans2022}.

The questions above motivate our present work. In order to generate a statistically complete census of Milky Way- and LMC-mass objects in a cosmological context, we make use of a $\left(100\,{\rm Mpc}\right)^3$ periodic dark matter-only simulation. We then augment this volume with a physically-motivated model of galaxy formation, generating a realistic population of galaxies (as predicted by the $\Lambda$ Cold Dark Matter model) across cosmic time. Finally, we construct galaxy merger trees (built on the merger trees of the haloes and subhaloes) in order to track the time evolution and eventual fate of the LMC-mass galaxies of interest. 

This paper is organised as follows. In Section~\ref{sec:numerics} we describe the simulations and the semi-analytic model of galaxy formation used in this work, as well as lay out the criteria we use to identify the potential progenitors of stellar haloes in our galaxy catalogues. Our main results are presented in Section~\ref{sec:results}, in which we contrast the properties of these galaxies with analogues at the same mass that are not destroyed, but rather survive until the present day. Finally, Section~\ref{sec:conclusions} provides a summary.

\section{Numerics}
\label{sec:numerics}

In this section, we describe the set of numerical simulations used in this work (Section~\ref{sec:simulations}) as well as the semi-analytic model of galaxy formation, \galform{}, that is used to populate dark matter haloes identified in our simulation with a realistic galaxy population (Section~\ref{sec:galform}).

\subsection{Cosmological simulations}
\label{sec:simulations}

We make use of the {\it Copernicus complexio Low Resolution} (\Color{}) simulation, a cosmological $N$-body simulation presented in \cite{Hellwing2016} and \cite{Bose2016}. This is a dark matter-only simulation performed in a periodic box of side length 100 Mpc using 1620$^3$ dark matter particles, corresponding to a mass resolution of $m_p = 8.8 \times 10^6\,{\rm M}_\odot$ per particle. The initial conditions for this run were set after assuming cosmological parameters derived from WMAP-7 data \citep{Komatsu2011}, with: $\Omega_m = 0.272$, $\Omega_\Lambda = 0.728$ and $h = 0.704$, where $h$ is related to the present-day Hubble constant, $H_0$, by $h = H_0/100{\rm kms}^{-1}{\rm Mpc}^{-1}$. The spectral index of the primordial power spectrum is $n_s = 0.967$, and the linear power spectrum is normalised at $z=0$ taking $\sigma_8 = 0.81$. We evolve the simulation from $z=127$ until the present day using the the \PGadget{} code \citep{Springel2001,Springel2005}.

At each snapshot output during the course of the simulation, we first identify candidate dark matter haloes using a friends-of-friends algorithm \citep{Davis1985}, which connects dark matter particles separated by at most 20 per cent of the mean interparticle separation, and subsequently identify bound structures within these groups using the \Subfind{} algorithm \citep{Springel2001b}. \Subfind{} identifies subsets of the particle distribution that are bound gravitationally, and may therefore be plausible sites for galaxies to form. We retain only \Subfind{} structures containing at least 20 bound particles, thereby yielding a minimum `resolved' halo mass of $1.8\times10^8\,{\rm M}_\odot$ in \Color{}. Throughout this paper, we define the physical extent of a dark matter halo by the radius $r_{200}$, which is the radius within which the mean density of the halo is 200 times the critical density of the Universe. Unless specified otherwise, the mass of haloes is quoted in terms of $M_{200}$, the total mass in dark matter particles enclosed within $r_{200}$. 

Augmenting a population of dark matter haloes with semi-analytic galaxies requires knowledge of their growth histories, and a census of important merger events that shape these histories. To this end, we use the substructures identified by \Subfind{} to serve as the roots for building merger trees. Associations between (sub)haloes in subsequent output times are established by identifying pairs of objects that share some fraction of their most-bound particles from one output time to the next. The method is described in detail in \cite{Jiang2014}. The (sub)halo merger trees are then traversed to generate galaxy populations using the \galform{} semi-analytic model of galaxy formation, which we describe in the following subsection.

\subsection{Semi-analytic galaxy formation}
\label{sec:galform}

We use the \galform{} semi-analytic model of galaxy formation to add galaxies to the dark matter (sub)haloes identified in the \Color{} simulation. A semi-analytic model provides a computationally cheap (relative to hydrodynamical simulations) and methodologically flexible way to generate a synthetic galaxy population to a dark matter-only simulation in post-processing \citep[e.g.][]{Kauffmann1993,Cole1994,Somerville1999,Cole2000,Croton2006,Benson2012,Henriques2015,Lagos2018}.

\galform{}, first presented in \cite{Cole1994} and \cite{Cole2000} follows the properties of subhalo merger trees and populates them with galaxies by solving coupled differential equations that encapsulate the physics of gas cooling in haloes, star formation, feedback from stars and black holes, the evolution of stellar populations, and chemical enrichment and recycling. In this paper we make use of the version of \galform{} described in \cite{Lacey2016} which combines several features from previous versions of the model, such as the inclusion of a top-heavy initial mass function (IMF) in starbursts, which is required to reproduce the abundance of star-forming sub-millimetre galaxies \citep{Baugh2005}; the model of feedback from active galactic nuclei introduced by \cite{Bower2006} which regulates the growth of massive galaxies; and a star formation law that depends on the molecular gas abundance within the interstellar medium \citep{Lagos2011}. \galform{} also makes predictions for the broad-band luminosities of its synthetic galaxies using the stellar population synthesis model of \cite{Maraston2005}. 

The free parameters in the \cite{Lacey2016} model are calibrated in order to reproduce existing constraints on the (present day) optical and near-infrared luminosity functions, the black hole-bulge mass relation, the HI mass function and the fraction of early- and late-type galaxies. The \galform{} model also includes prescriptions for modelling cosmic reionisation and its effects on the star forming capacity of galaxies; it also includes a methodology for tracking galaxies formed in subhaloes that are disrupted below the resolution limit of the simulation (so-called `orphan galaxies'). However, the mass scales where these are relevant corresponds to the regime of the ultrafaints, which we do not consider in this work. For details about how these processes are modelled, we refer the reader to \cite{Bose2020}.

In what follows, we will refer to a `central' galaxy as a \galform{} galaxy that has formed within the largest subhalo of a friends-of-friends group. `Satellites' of this central galaxy are defined as \galform{} galaxies that are located within a distance of $r_{200}$ from the central.

\subsection{Selecting progenitors of the stellar halo}
\label{sec:selection}

Our primary goal in this work is to track the properties of galaxies that could feasibly act as progenitors of stellar haloes in Milky Way-mass galaxies. To this end, we define a Milky Way mass galaxy as being:
\begin{enumerate}
    \item A {\it central} galaxy at $z=0$, with
    \item Stellar mass in the range $10 \leq \log\left[ M_\star/{\rm M}_\odot\right] \leq 10.7$, and
    \item Hosted in dark matter haloes in the mass range $11.8~\leq~\log\left[ M_{200}/{\rm M}_\odot\right]~\leq 12.3$.
\end{enumerate}
Furthermore, we define a candidate stellar halo progenitor as being:
\begin{enumerate}
    \item A {\it central} galaxy at $z=2$, with
    \item Stellar mass in the range $7.9 \leq \log\left[M_\star/{\rm M}_\odot \right]\leq9.3$, which
    \item Becomes a {\rm satellite} (i.e., is accreted into the virial radius) of a larger host galaxy between $1\leq z \leq2$, and whose
    \item Merger tree points to a Milky Way-mass galaxy (as defined above) as its descendant galaxy at $z=0$.
\end{enumerate}
These conditions guarantee that the sample of interest consists of objects that have merged onto a central, Milky Way-mass galaxy by the present day (i.e., they are ``destroyed'') but were LMC-mass objects at the time of their accretion 8-10 Gyr ago i.e., such that they became part of larger systems between $1\leq z \leq2$. Note that \galform{} does not track a separate stellar halo component; the stellar mass of an accreted satellite that is destroyed is just assigned to the stellar mass of the central galaxy of the host halo.
\begin{table}
    \centering
    \begin{tabular}{c|c}
    \hline
    Type & Number  \\
    \hline \hline
    Milky Way-mass analogues at $z=0$ & 329 \\
    whose main progenitors were LMC-mass at $z=2$ & 146 \\
    \hline
    LMC-mass analogues at $z=2$ & 16,999 \\
    \,\,\,which become satellites by $z=1$ & 5082\\
    and are hosted by Milky Way-mass haloes & 94\\
    \,\,\,and are destroyed by $z=0$ & {\bf 91}\\
    \hline
    \end{tabular}
    \caption{Demographics of the galaxy populations of interest in this work as extracted from the \Color{} simulation. For descriptions of how each galaxy set is defined, we refer the reader to the discussion in Section~\ref{sec:selection}. The number in bold corresponds to the population of ``destroyed'' galaxies that we use as proxies for progenitors of the stellar haloes of Milky Way-mass galaxies. Note that condition in the third row (number of objects that are satellites by $z=1$) searches for satellite galaxies across hosts of {\it all} masses, and not just Milky Way-like hosts (the number is 94 in the latter instance).}
    \label{tab:obj_counts}
\end{table}

\begin{figure}
\centering
\includegraphics[width=\columnwidth]{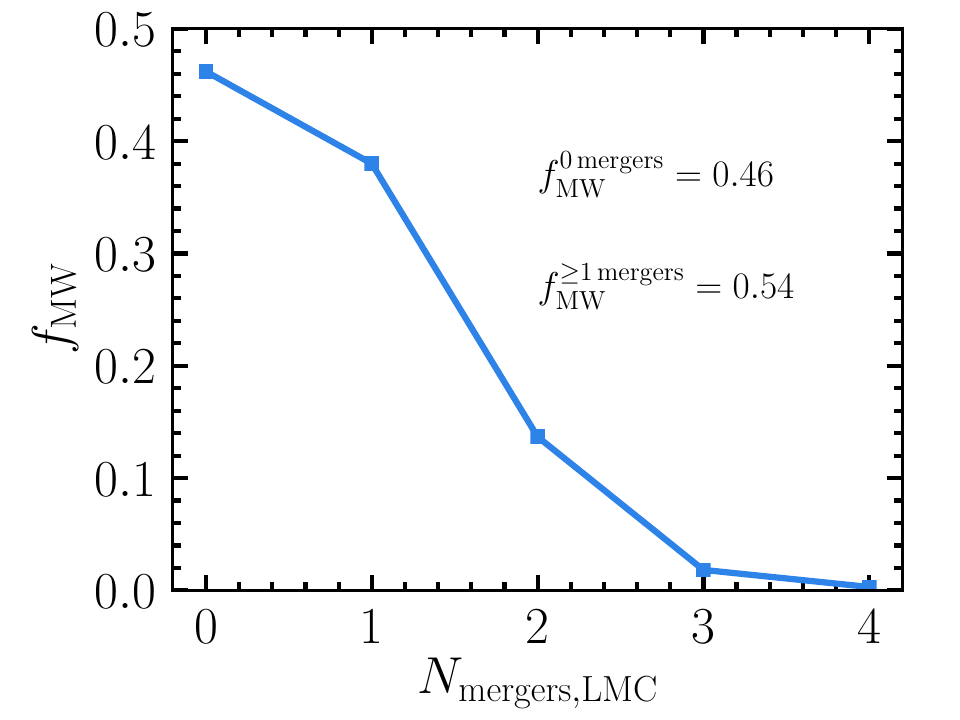}
\caption{The fraction of Milky Way-mass hosts, $f_{{\rm MW}}$, that have accreted a certain number of LMC-mass galaxies, $N_{{\rm mergers,LMC}}$, since $z=2$ that are ultimately destroyed by $z=0$. Note that in this figure, we consider destroyed LMC-mass galaxies that have been accreted at {\it any} point in the past i.e., they do not have to be identified as satellites at $z=1$. Just over half the Milky Way-mass hosts contain one or more destroyed LMC-mass galaxies, while just under half either never accrete such an object or, if they do, that satellite survives until the present day.}
\label{fig:Nmerge_MW}
\end{figure}

We summarise the total number of galaxies identified using these classifications in Table~\ref{tab:obj_counts}. We adopt a somewhat generous range for the stellar mass of the progenitor galaxy at $z=2$ in order to boost the statistics of our sample. While we identify nearly 17,000 galaxies at $z=2$ in the stellar mass range appropriate to an LMC-like progenitor, only 91 ($\approx 0.5\%$) galaxies fulfil the full set of criteria that we use to determine a potential progenitor of a Milky Way stellar halo. In what follows, we will refer to this smaller set of galaxies as the subset of `destroyed' galaxies; the remainder of the set will be referred to as `surviving' galaxies. Interestingly, of the 5082 LMC-mass galaxies that are identified as satellites at $z=1$, only 94 of these are satellites of Milky Way-mass galaxies (or, more specifically, their progenitors at $z=1$); the remainder end up as satellites of systems more massive than the mass range used to identify Milky Ways. 91 out of those 94 satellites are destroyed between $z=1$ and $z=0$, showing that LMC-mass satellites that are accreted relatively early on are not expected to survive until the present day. This is consistent with estimates of the infall redshift of the LMC itself, which is estimated to have been accreted into the halo of the Milky Way relatively recently \citep[see, e.g.,][]{Besla2015, Shao2018, Evans2020}. Finally, we also note that 146 out of 329 Milky Way-mass analogues at $z=0$ ($\approx44\%$) had main progenitors that were LMC-mass analogues at $z=2$.

The statistics listed in Table~\ref{tab:obj_counts} make it clear that destroyed LMCs (based on the criteria we have set out earlier in the section) are relatively infrequent. It is worth considering how often a Milky Way-mass host accretes (and subsequently destroys) an LMC-mass galaxy {\it at all}. In other words, we can remove condition (iii) in the list of criteria defining a stellar halo progenitor and consider LMC-mass galaxies that become satellites at {\it any} point after $z=2$.  Figure~\ref{fig:Nmerge_MW} shows the fraction of Milky Way-mass hosts at $z=0$, $f_{{\rm MW}}$, that have experienced a certain number of LMC-mass mergers, $N_{{\rm mergers,LMC}}$, since $z=2$. We find that almost half the Milky Way sample ($\sim 46\%$ of 329 total hosts) have {\it never} experienced a merger of this kind (or, if it has, the satellite is not destroyed by $z=0$\footnote{We find that 83 out of our 329 Milky Way-mass hosts ($\sim25\%$) contain an {\it undestroyed} satellite of this kind at $z=0$. This corresponds to a total of 96 satellite galaxies at the present day.}). Of the remainder that do experience such events, 38\% have accreted 1 satellite; 13\% have accreted 2 satellites; 1.7\% have accreted 3 satellites; 0.3\% (i.e., one Milky Way) has accreted 4 such satellites since $z=2$.

\begin{figure}
\centering
\includegraphics[width=\columnwidth]{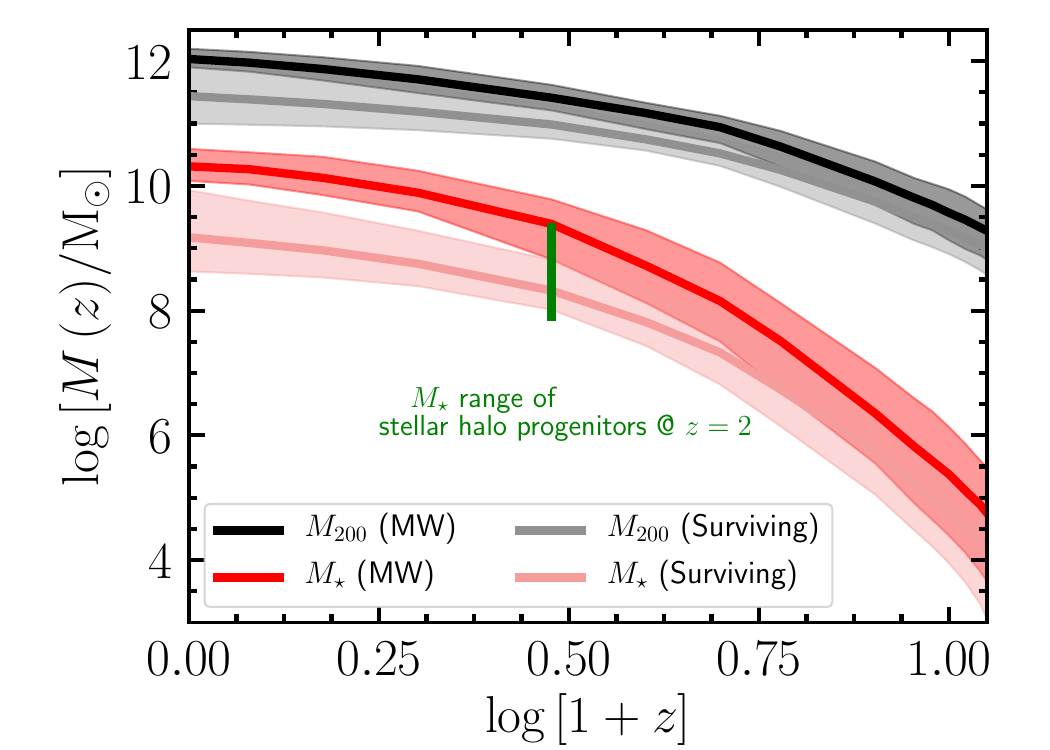}
\caption{The halo (black) and stellar (red) mass growth histories of the Milky Way-mass (darker shade) and LMC-mass analogues (lighter shade). The vertical green line shows the stellar mass range (at $z=2$) that is used to select our candidate LMC-mass stellar halo progenitors. The shaded regions mark the 68\% scatter around the median (shown as the solid lines). The relatively wide mass bin we adopt results in some overlap with objects that are more consistent with being the main progenitor of a Milky Way-mass galaxy at $z=0$. Any galaxy identified as such is removed from our list of surviving/destroyed LMC-mass galaxies. Note that the lighter shade red curves extend to $z=0$ for just the surviving LMCs.}
\label{fig:MAH}
\end{figure}

\begin{figure*}
    \centering
    \includegraphics[width=\textwidth]{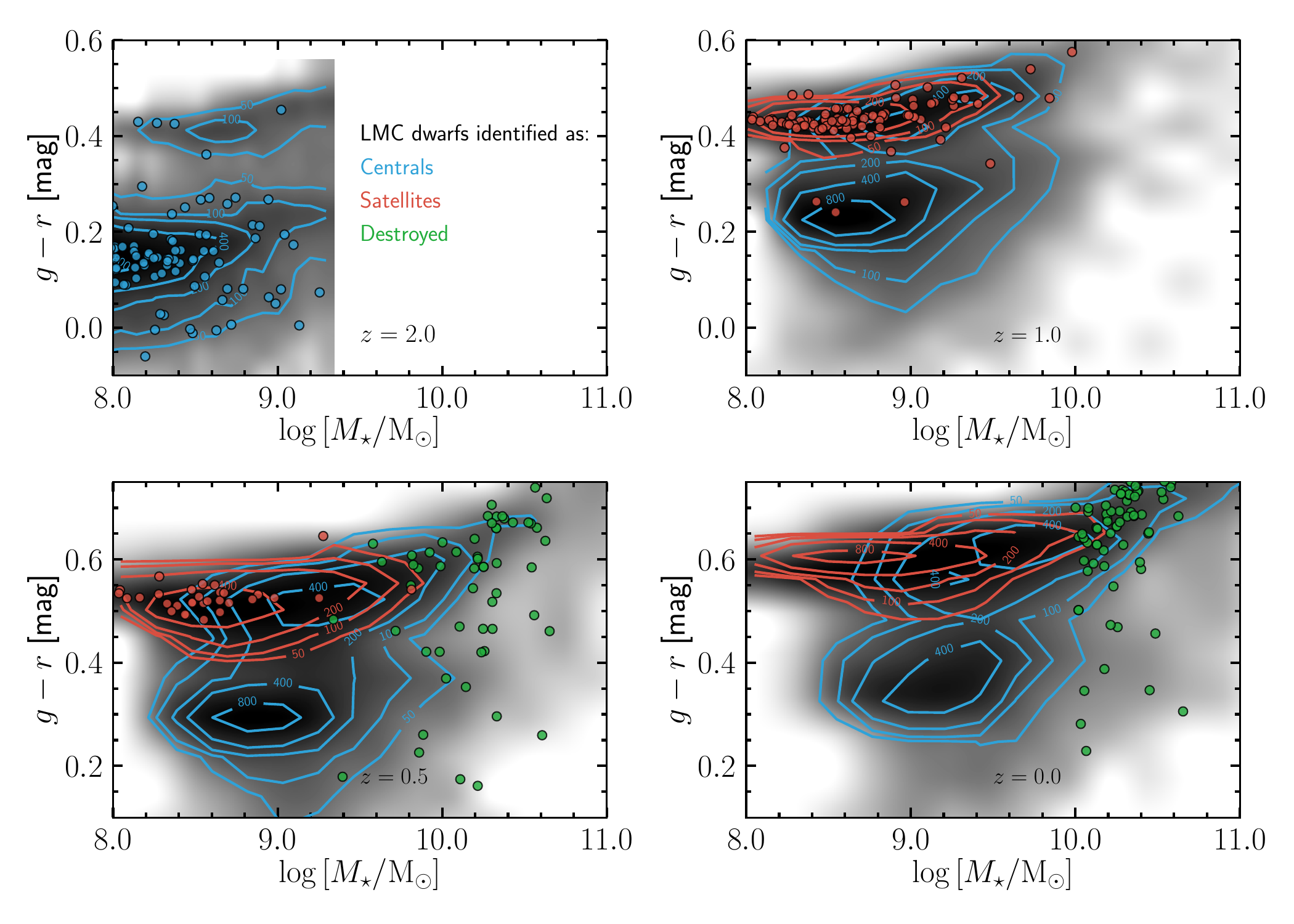}
    \caption{Evolution of galaxies in the plane of colour and stellar mass between $z=2$ and $z=0$. The grey histogram shows the abundance of all  galaxies identified in the mass range $7.9\leq\log\left[M_\star/{\rm M}_\odot\right]\leq9.3$ at $z=2$. The coloured dots represent `destroyed galaxies': those objects that are identified as candidate progenitors of the stellar halo for Milky Way-like galaxies at $z=0$ (see main text for the full list of criteria used to distinguish this population). Blue and red colours, respectively, categorise central and satellite galaxies at each output time, with their corresponding counts displayed with contours. Once an accreted dwarf is destroyed, its stellar mass is added to that of the central galaxy of the host Milky Way mass halo; we mark these `centrals' with green dots.}
    \label{fig:mstar_vs_colour}
\end{figure*}

\begin{figure*}
    \centering
    \includegraphics[width=\textwidth]{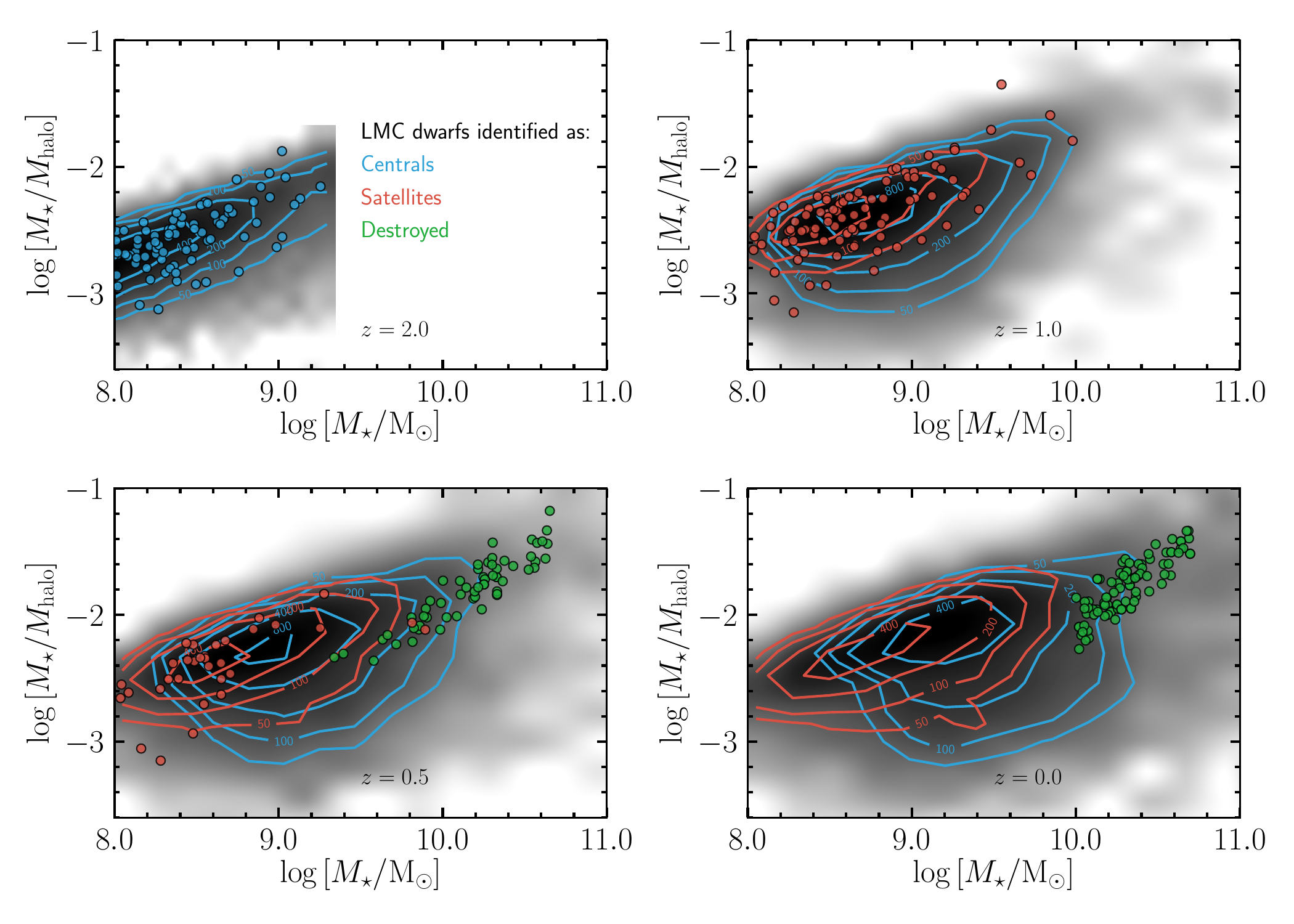}
    \caption{The evolution of galaxies in the plane of $M_\star$ vs. $M_\star/M_{{\rm halo}}$. For central galaxies, $M_{{\rm halo}}$ is just the $M_{200}$ of the friends-of-friends group at the current time; for satellites, $M_{{\rm halo}}$ is the mass of the halo when it was last identified as a central (i.e., at the time of infall). Symbol styles and colours are identical to those in Fig.~\ref{fig:mstar_vs_colour}.}
    \label{fig:mstar_vs_mhalo}
\end{figure*}

Figure~\ref{fig:MAH} shows the redshift evolution of stellar (red) and halo (black) mass for our sample of Milky Way-mass and `surviving' progenitor galaxies (darker and lighter shades, respectively). Since the two sets of galaxies survive until the present day, both show sustained growth in both stellar and halo mass with the rate of stellar mass growth slowing down from around $z=2$. While the median halo mass at $z=0$ differs by 0.5 dex, the median stellar mass shows a larger disparity, differing by more than an order of magnitude. This difference in the relative growth of the stellar and dark matter components of the two populations of objects is simply a reflection of the slope of the stellar-to-halo mass relation in this regime. This is discussed in more detail in Section~\ref{sec:stellar_halo_mass} and Figure~\ref{fig:mstar_vs_mhalo}. The set of `destroyed' galaxies tracks the evolution of the `surviving' set until these galaxies become satellites at $z=1$; thereafter, they no longer grow in stellar or halo mass. This will be discussed in more detail in Section~\ref{sec:stellar_halo_mass}. Note that a `surviving' galaxy at $z=0$ may be a central or a satellite, although  centrals overwhelmingly dominate the population.

The comparison of the mass growth histories of the destroyed and surviving galaxies suggests a divergence in the evolutionary pathways of these two categories. Both sets of galaxies are descendants of progenitors of the same mass at high redshift (until $z=1$); whereas the survivors continue to grow in mass, the objects that are accreted into the dark matter haloes of Milky Way-mass galaxies halt further stellar mass growth until their eventual disruption forming stellar haloes. In this sense, the progenitors of the stellar halo of Milky Way-mass galaxies may themselves be thought of as `failed' Milky Ways -- i.e., were they not accreted into a bigger galaxy, their eventual fate would not be dissimilar to our own Galaxy. In fact, the main progenitors of $\sim45\%$ of the Milky Way-mass galaxies at $z=0$ (146 out of 329, see Table~\ref{tab:obj_counts}) fall in the LMC analogue range at $z=2$. We examine this further in the following section, where we analyse the properties of the destroyed galaxies in more detail. 

\section{Results}
\label{sec:results}

Having outlined the criteria used to define candidate progenitors of stellar haloes in Milky Way-mass galaxies, we now present our main results regarding their observable properties, and the evolution of these properties from $z=2$ until the present day. In particular, we consider their evolution in the plane of colour-stellar mass and the stellar-to-halo-mass relation with respect to the overall galaxy population. We also consider the differences in the properties, if any, between those galaxies that end up being `destroyed' and those that survive and grow until $z=0$. 

\subsection{The evolution of galaxy colours}
\label{sec:colour}

We begin by considering the distribution of galaxy colours for the `destroyed' galaxy set (as defined in Section~\ref{sec:selection}) compared to the overall galaxy population. Figure~\ref{fig:mstar_vs_colour} shows the distribution of  ($g-r$) colours as a function of stellar mass, $M_\star$, measured in the galaxy catalogues at $z=2$ (top left), $z=1$ (top right), $z=0.5$ (bottom left), and $z=0$ (bottom right). The $g-r$ colours at each redshift have been converted from the rest-frame of the galaxy to the observer's frame and have been corrected for extinction by dust using the methodology described in \cite{Lacey2016} (Section 3.9.2). Galaxies identified as centrals are shown in blue, while those identified as satellites are marked in red. 

The grey histogram in the background represents the population of {\it all} central galaxies identified at $z=2$ that are identified in the mass range $7.9\leq\log\left[M_\star/{\rm M}_\odot\right]\leq9.3$, while the circles correspond to the subset of these objects that are destroyed by $z=0$; by definition, all of these galaxies are identified as centrals initially. The familiar bimodal distribution of galaxy colours is immediately evident. It is interesting to note that the set of galaxies that are ultimately destroyed (which, we remind, are the ones considered to be potential progenitors of the stellar haloes of Milky Way-mass hosts) span the full range of colours, albeit with them predominantly ($> 90\%$) being skewed towards bluer colours. 
\begin{figure*}
    \centering
    \includegraphics[width=0.45\textwidth]{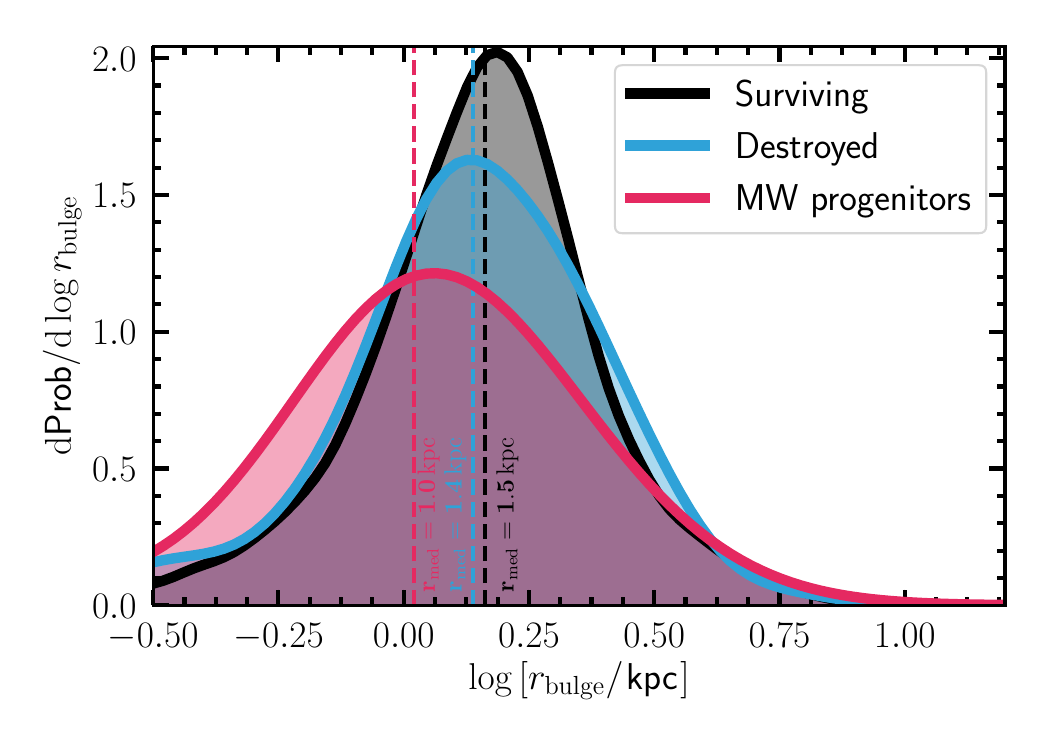}
    \includegraphics[width=0.45\textwidth]{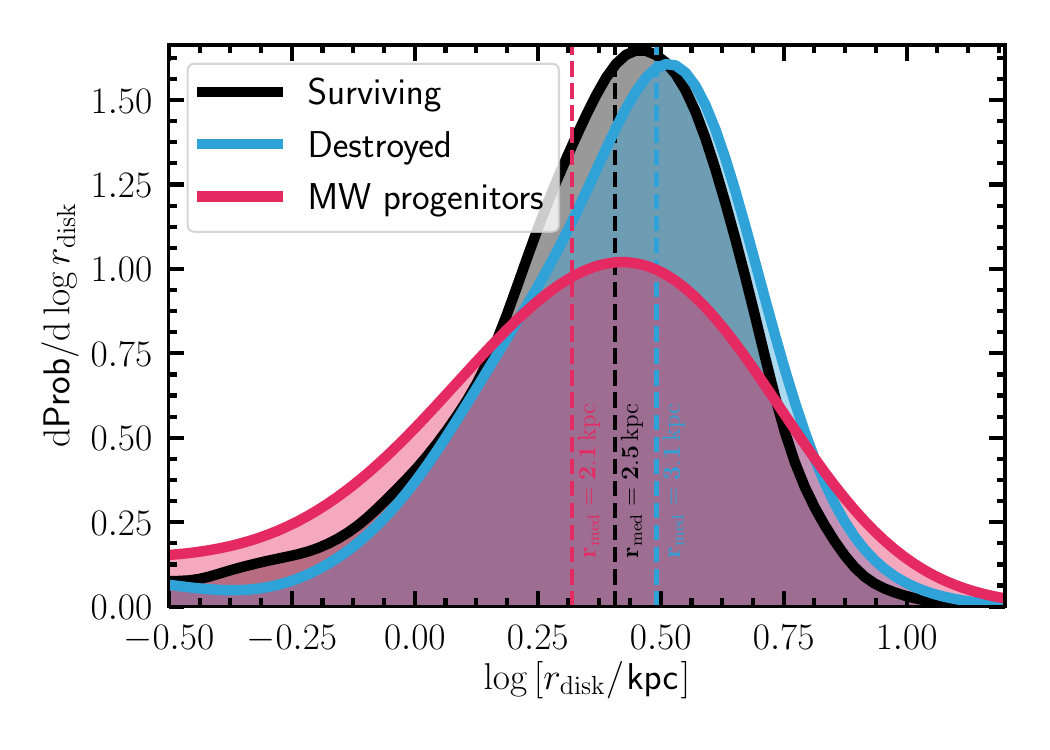}\\
    \includegraphics[width=0.45\textwidth]{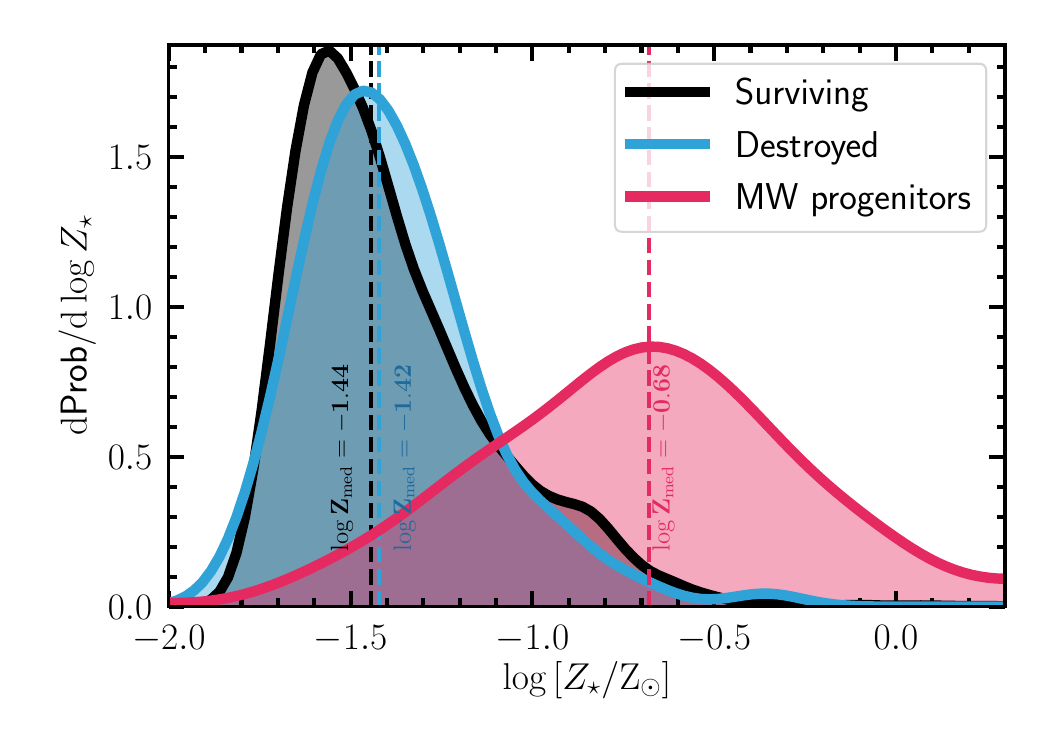}
    \includegraphics[width=0.45\textwidth]{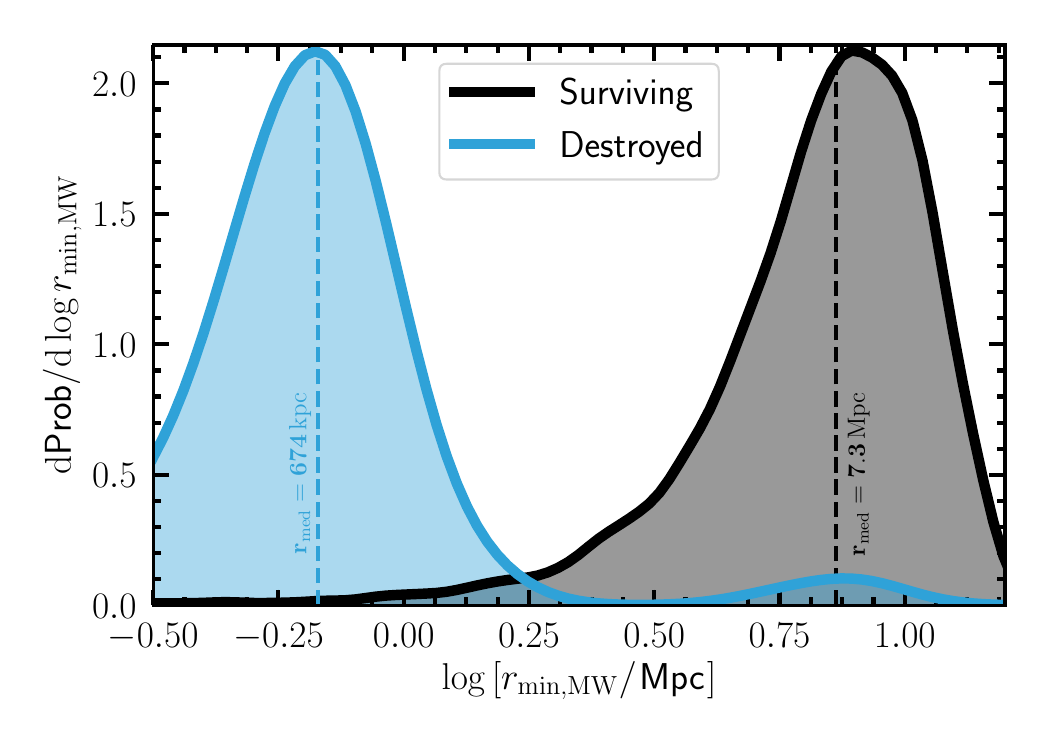}
    \caption{A selection of galaxy properties measured at $z=2$. The quantities we consider are: bulge size, $r_{{\rm bulge}}$; disk size, $r_{{\rm disk}}$; metallicity, $Z_\star$; and distance to the nearest progenitor of a Milky Way-like galaxy, $r_{{\rm min,MW}}$. The black histogram shows the distribution of these properties for galaxies in the mass range $7.9\leq\log\left[M_\star/{\rm M}_\odot\right]\leq9.3$ at $z=2$ that survive until present-day; the histogram in blue shows galaxies in that mass range that are destroyed before $z=0$ (i.e., potential stellar halo progenitors). Finally, the histogram in red shows the properties of galaxies identified as progenitors of Milky Way-like hosts. The dashed vertical lines indicate the median value of each quantity for the corresponding galaxy population.}
    \label{fig:prog_props}
\end{figure*}

By $z=1$ each of these galaxies have fallen into larger haloes (which themselves grow to the mass range corresponding to the Milky Way at by $z=0$) and are therefore now labelled as satellites. Star formation in these objects now largely shuts off resulting in the vast majority of these galaxies now appearing red, although a small portion ($\lesssim5\%$) are still located among the bluer population. Indeed, we find that the specific star formation rate of these galaxies $\log\left[ {\rm sSFR} / {\rm Gyr}^{-1} \right] \leq 1.0$, which is shorter than the typical mass doubling timescale of galaxies at these redshifts \citep[e.g.][]{Feulner2005}. 

As the galaxy population evolves to $z=0.5$ (bottom left panel in Figure~\ref{fig:mstar_vs_colour}), we once again find that the set of progenitor galaxies we have been tracking since $z=2$ are split between centrals and satellites. The ones that are marked as centrals once again correspond to galaxies that have been completely destroyed after infall; their descendant in the merger tree is now the central galaxy of the host halo that the erstwhile satellite fell into. We mark these `centrals' in green to indicate that these are destroyed galaxies that have merged with the central galaxy of the Milky Way host. We find that this is true for more than half the population ($\sim 65\%$) we have tracked until now. If these destroyed galaxies are indeed ones that dominate the mass of the stellar haloes of Milky Way-like galaxies, our model predicts that the bulk of the stellar halo may therefore already be in place by $z=0.5$. On the other hand, the LMC-mass progenitors are still identifiable as satellites in a third of the Milky Way-mass hosts (i.e., they have not yet been destroyed by the tidal forces or by merging on to the central galaxy) suggesting that the stellar halo in these hosts only assemble relatively recently ($z\leq0.5$).

\subsection{The evolution of the stellar-to-halo mass relation}
\label{sec:stellar_halo_mass}

Next, we consider the range of halo masses hosting galaxies that are potentially the progenitors of stellar haloes, and compare their distribution their counterparts that survive. In Figure~\ref{fig:mstar_vs_mhalo} we consider the stellar-to-halo mass evolution of the destroyed LMC-mass progenitors from $z=2$ to $z=0$. The colour scheme chosen is identical to that in Figure~\ref{fig:mstar_vs_colour}. The grey histogram and coloured contours, once again, show the distribution of all central galaxies in the mass range $7.9\leq\log\left[M_\star/{\rm M}_\odot\right]\leq9.3$ that were identified at $z=2$.  

We see that the destroyed galaxies are distributed in much the same way as the overall galaxy population in this mass range; there is no obvious systematic tendency for them to live in over- or under-massive dark matter haloes given their stellar mass. At $z=2$, the galaxies we track are typically hosted in haloes with mass $10.5~\leq~\log\left[M_{200}/{\rm M}_\odot\right]~\leq~12.0$. Once they transition to being satellites from $z=1$ to $z=0.5$, these galaxies are frozen in position in this diagram due to the fact that there is no further growth in stellar mass (apart from any residual star formation that takes place in the cold gas reservoir of these galaxies), while the halo mass also does not change. The exceptions to this are the galaxies that are already destroyed by $z=0.5$, with their stellar and halo mass now assigned to the stellar and halo mass of the central galaxy of the Milky Way host. Finally, by $z=0$ all destroyed galaxies are clustered on the right hand side of the diagram as all satellites have by now merged on to the central, by definition. In addition to masses, we have also checked that, on average, the formation times of haloes (defined as the epoch at which they reach half their final mass) is very similar between these populations of galaxies. 

It is clear to see from Figures~\ref{fig:mstar_vs_colour}~\&~\ref{fig:mstar_vs_mhalo} that the evolution of galaxies from $z=2$ to $z=0$ is very similar between stellar halo progenitor galaxies and the remainder of the galaxy population in the corresponding mass range. In the following subsection, we will consider the possibility of identifying stellar halo progenitors before they are accreted onto a more massive host system.

\subsection{The observed properties of stellar halo progenitors at \texorpdfstring{$z=2$}{}}
\label{sec:observed_distr}

We now consider the observational properties of stellar halo progenitor galaxies at $z=2$. Our aim is to identify what differences, if any, there are between the galaxies at $z=2$ (i.e., before they become satellites) that contribute to the stellar haloes of Milky Way-mass galaxies, and the surviving population that includes the main progenitors of the Milky Way themselves.  

Figure~\ref{fig:prog_props} shows probability distributions of a selection of galaxy properties for the destroyed (blue), surviving (black), and Milky Way progenitor (red) populations. In particular, we consider distributions of the bulge size, $r_{{\rm bulge}}$ (top left), disk size, $r_{{\rm disk}}$ (top right), and stellar metallicity, $Z_\star$ (bottom left), with each panel showing the predictions from \galform{} at $z=2$. The probability distributions shown are based on kernel density estimates from the \galform{} data for each population. 

It becomes immediately evident from Figure~\ref{fig:prog_props} that the distribution of the $z=2$ galaxy properties is very similar for both the surviving and the destroyed galaxies, both in the median and in the scatter. At $z=2$, the median bulge and disk sizes, respectively, are around 1.5~kpc and $\sim$3~kpc. Interestingly, the model suggests that some disks as large as 6 kpc or bigger may already be in place at this time. The distribution of stellar metallicities (as shown in the bottom left corner of Figure~\ref{fig:prog_props}) is nearly identical for the surviving and destroyed galaxies, with the distribution peaking around $\log\left[Z_\star/Z_\odot\right]\approx-1.4$.

The main progenitors of Milky Way-mass galaxies are somewhat distinct in their distribution of galaxy properties at this time. Most notably, the PDFs for the bulge and disk sizes are broader than for the LMC-mass progenitors and in fact the the median bulge size in the Milky Way progenitor case is shifted to slightly lower values ($\approx1$~kpc), although still statistically consistent with the destroyed and surviving populations. More clear differences are noted in the distribution of stellar metallicities, where apart from the larger scatter, the distribution also peaks at much larger metallicity ($\log\left[Z_\star/Z_\odot\right]\approx-0.7$) in the Milky Way progenitor population. This is not unexpected given the stellar mass-metallicity relationship of galaxies, in which entities with larger stellar mass are found to be more metal-rich. As we have seen in Figure~\ref{fig:MAH}, the main progenitors of Milky Way-mass galaxies are typically an order of magnitude more massive than the LMC-mass analogues at $z=2$. That said, in the narrow range of stellar masses when the two populations overlap, the properties are indeed very similar.

Seeing as the observable properties of surviving and destroyed galaxies at $z=2$ are nearly indistinguishable, it raises the question of what ultimately determines whether or not a $z=2$ galaxy ends up as a progenitor of a bigger galaxy or survives until the present day. Perhaps the most obvious factor is the proximity of each galaxy to its nearest Milky Way-mass progenitor at $z=2$. We quantify this in the bottom right panel of Figure~\ref{fig:prog_props}, in which we compare histograms of the mean distance to the nearest Milky Way-like galaxy, $r_{{\rm min, MW}}$, between the destroyed and surviving populations. We now see a stark contrast between these two populations. The median distance at $z=2$ for the surviving galaxies is of order 7~Mpc, while that for the destroyed galaxies is just $\approx$ 670~kpc -- within a factor of a few of the mean virial radius of Milky Way-mass host haloes at these redshifts. The near total exclusion of the two histograms is interesting to note. All but two of the destroyed galaxies are located within 1.5 Mpc of their nearest Milky Way-mass neighbour; on the other hand, more than 97\% of the surviving galaxies are identified at distances greater than 1.5 Mpc from their nearest Milky Way-mass neighbours.

In addition to the properties examined in this section, we have also compared the total gas mass, gas metallicity, and star formation rates of the surviving and destroyed galaxies at $z=2$. Our conclusions are consistent with what we have observed above: that there are no demonstrably different properties that distinguish one galaxy population from the other. According to our model, it is only the proximity to a neighbouring massive galaxy that determines the future evolution of these galaxies. Given the relatively small separation between the LMC-mass progenitor and the main progenitor of the Milky Way, it is likely that a survey that searches for these objects at $z=2$ will find them in the same field-of-view \citep[see, e.g., ][]{Evans2022}. 

\section{Discussion \& Conclusions}
\label{sec:conclusions}

We have presented an analysis of the properties of galaxies at $z=2$ with mass comparable to that of the LMC by segregating this population into two categories: those that survive until the present day, and the (smaller) set of galaxies that are destroyed before $z=0$ within the host halo of an Milky Way-mass galaxy. The latter category is especially interesting as this mass range is expected to dominate the stellar haloes of Milky Way-like galaxies, which are formed through the accretion and subsequent destruction of dwarf galaxies. To this end, we make use of the \galform{} semi-analytic model of galaxy formation, run on the $\left(100\,{\rm Mpc}\right)^3$ {\it Copernicus complexio Low Resolution} (\Color{}) dark matter-only simulation to select the populations of interest based on the criteria listed in Section~\ref{sec:selection}. Our main findings are as follows:
\begin{enumerate}
    \item While we find several candidate galaxies ($\sim$17,000) that are in the appropriate mass range to be LMC analogues, only a very small fraction of these ($\approx0.5\%$) satisfy the full set of conditions we set out to identify potential stellar halo progenitors (Table~\ref{tab:obj_counts}). In the vast majority of cases when an LMC-mass dwarf has been accreted by $z=1$, this satellite is destroyed by $z=0$. Around 46\% of the Milky Way-mass hosts in our sample contain no destroyed LMC-mass galaxies at all (Figure~\ref{fig:Nmerge_MW}).
    \item On the other hand, the LMC-mass dwarfs that do not become satellites and survive until present day continue to grow in stellar and dark matter mass. At $z=0$, their median halo and stellar masses, respectively, are 0.5 dex and 1.0 dex smaller than an average Milky Way-mass galaxy (Figure~\ref{fig:MAH}). 
    \item We then investigate the evolution of the destroyed and surviving galaxies in the colour-stellar mass diagram. At $z=2$, the majority of galaxies are identified as blue, star-forming galaxies; once they are identified as satellites at $z=1$, they transition to redder $g-r$ colours with specific star formation rates that are consistent with them being quenched (Figure~\ref{fig:mstar_vs_colour}). This is an outcome of the \galform{} model in which gas accretion is shut off as a galaxy becomes a satellite. A handful  of satellites are still identified as blue. 
    \item Around $65\%$ of the stellar halo progenitor galaxies are destroyed by $z=0.5$, suggesting that the bulk of the stellar halo mass in Milky Way-mass galaxies may already be in place by this time. 
    \item The distribution of halo masses hosting stellar halo progenitors is consistent with that of the surviving population. We find that at $z=2$, these galaxies are found in haloes with mass $10.5~\leq~\left[M_{200}/{\rm M}_\odot\right]~\leq~12.0$ (Figure~\ref{fig:mstar_vs_mhalo}).
    \item Finally, we investigate a variety of observable galaxy properties at $z=2$ in an attempt to identify characteristics of destroyed galaxies that may distinguish them from the rest of the population in a targeted survey. In general, we find that these galaxies are fully representative of typical LMC-mass analogues at $z=2$, and we do not find any unique observational characteristics distinguishing these galaxies (Figure~\ref{fig:prog_props}). The only marked difference between a destroyed galaxy and one that survives until present day is their proximity to a neighbouring object that is a progenitor of a Milky Way-mass galaxy. At $z=2$, the median distance between a surviving galaxy and its closest Milky Way progenitor is around 7 Mpc, whereas this distance reduces to around 670 kpc for destroyed galaxies. This difference in the immediate environments of the two sets of galaxies is ultimately all that determines the divergent evolutionary pathways of LMC-mass galaxies after $z=2$. 
\end{enumerate}
One can draw some interesting conclusions from our observations. Despite the fairly generous mass range we have adopted for identifying potential stellar halo progenitor galaxies, we find that a very small number of these objects (91, $\approx 0.5\%$ of the total population) satisfy the full set of conditions we have defined in Section~\ref{sec:selection} to mark destroyed galaxies. Even if we were to drop the timing condition (iii) and simply count {\it all} LMC-mass galaxies that are destroyed in a Milky Way-mass host by $z=0$, this number increases from 91 to 237, or $\approx1.4\%$ of the full population of LMC-mass galaxies at $z=0$. As shown in Figure~\ref{fig:Nmerge_MW}, just over half the Milky Way-mass hosts in our sample contain one or more destroyed LMCs at all (i.e. accreted at any time since $z=2$). If the formation of stellar haloes in Milky Way-mass galaxies is ubiquitous, this suggests that their assembly through the accretion of one or a few LMC-mass dwarf galaxies may not be the only formation channel \citep[see, e.g.,][for estimates of the number of dwarf galaxy mergers comprising stellar haloes in hydrodynamical simulations of Milky Ways]{Monachesi2019,Fattahi2020}.

It also interesting to note the near identical distribution of galaxy properties in the destroyed and surviving galaxy populations. In particular, the only feature (according to our model) that truly determines the trajectory of $z=2$ galaxies along one of these diverging pathways is its environment -- i.e., how close it happens to be to a neighbouring massive galaxy. A destroyed dwarf is simply unlucky enough to be accreted into a larger galaxy. As the surviving galaxies themselves follow evolutionary tracks not dissimilar to the Milky Way (albeit ending at lower final mass), the progenitor galaxies of stellar haloes can be considered to be `failed' Milky Ways. The similarity of their properties at $z=2$ indicates that the old stellar populations in the Milky Way today may be largely indistinguishable to the progenitors of the stellar halo in intrinsic properties like mass, metallicity etc. Due to differences in how they accumulate within the Milky Way, however, there may be substantial differences in their spatial and kinematic arrangement. The prospect of drawing this connection between high redshift progenitors and the present state of our Galaxy offers an exciting motivation for targeted surveys of galaxies at $z\gtrsim 1$.

\section*{Acknowledgements}

We are grateful to the anonymous referee for carefully reading this manuscript and suggesting changes that make it a more comprehensive study. SB is supported by the UK Research and Innovation (UKRI) Future Leaders Fellowship [grant number MR/V023381/1]. AD is supported by a Royal Society University Research Fellowship. AD acknowledges support from the Leverhulme Trust and the Science and Technology Facilities Council (STFC) [grant numbers
ST/P000541/1, ST/T000244/1]. AD thanks the staff at the Durham University Day Nursery who play a key role in enabling research like this to happen. This work used the DiRAC@Durham facility managed by the Institute for Computational Cosmology on behalf of the STFC DiRAC HPC Facility (\url{www.dirac.ac.uk}). The equipment
was funded by BEIS capital funding via STFC capital grants ST/K00042X/1, ST/P002293/1, ST/R002371/1 and ST/S002502/1, Durham University and STFC operations grant ST/R000832/1. DiRAC is part of the National e-Infrastructure.

\section*{Data availability}
The data used in this paper are available upon request to the corresponding author.




\bibliographystyle{mnras}
\bibliography{failedMW} 



\bsp	
\label{lastpage}
\end{document}